\documentclass[preprint,showpacs,amsmath,amssymb]{revtex4}

\usepackage{graphicx}
\usepackage{dcolumn}
\usepackage{bm}

\begin{document}

\title{Annealing-tuned Hall coefficient in single crystals of the YbNi$_2$B$_2$C heavy fermion}

\author{Sergey L. Bud'ko, and Paul C. Canfield}
\affiliation{Ames Laboratory US DOE and Department of Physics and Astronomy, Iowa State University, Ames, IA
50011, USA}

\date{\today}

\begin{abstract}
We present temperature-dependent magneto-transport measurements on as-grown and annealed YbNi$_2$B$_2$C single
crystals. Annealing causes drastic changes in the Hall coefficient, $R_H(T)$. Whereas for as-grown samples the
Hall coefficient is  negative between room temperature and 2 K, with a pronounced {\it minimum} at $\approx 22$ K,
for the samples annealed at $950^\circ$ C for 150 hours, $R_H(T)$ changes its sign twice in the same temperature
range: from negative to positive on cooling below $\sim 100$ K and back to negative below $\sim 10$ K, and has a
clear {\it maximum} at $\approx 45$ K. Intermediate temperature dependencies can be achieved by reducing the
annealing time. These findings are discussed within the framework of an annealing dependence of the skew
scattering in conjunction with the recent structural, thermodynamic and transport studies of the effects of
annealing in YbNi$_2$B$_2$C.
\end{abstract}


\maketitle

\section{Introduction}

YbNi$_2$B$_2$C is the heavy fermion \cite{yat96a,dha96a} member of a rich and complex family of quaternary
borocarbibides, RNi$_2$B$_2$C (R = rare earth). This family encompasses a wide range of physical phenomena:
interaction between local moment magnetism and superconductivity, peculiar, highly anisotropic, metamagnetism, and
strong electronic correlations \cite{can98a,mul01a}. Unique features of YbNi$_2$B$_2$C such as moderately high
electronic specific heat coefficient, $\gamma \approx 500$ mJ/mol K$^2$, well separated characteristic
temperatures/energy scales (no superconductivity or long-range magnetic order above 0.03 K, Kondo temperature,
$T_K \approx 10$ K and crystal electric field splitting, $T_{CEF} \approx 100$ K
\cite{yat96a,gra96a,ram00a,boo03a}), as well as its availability in a single-crystalline form, make it a model
system for detailed studies of strongly correlated Yb-based materials. Recently it was shown \cite{mia02a} that
whereas the effect of annealing on the resistivity of single crystals of non-magnetic (R = Y, Lu) and ''good local
moment'' (R = Gd-Tm) members of the RNi$_2$B$_2$C series is an ordinary decrease of the residual resistivity
consistent with Matthiessen's rule, for YbNi$_2$B$_2$C extraordinary annealing-induced changes were found in the
whole studied temperature range, 2 - 300 K. A subsequent detailed investigation \cite {avi02a} delineated the
effects of annealing on YbNi$_2$B$_2$C: thermodynamic properties (DC magnetic susceptibility, $\chi(T)$, and heat
capacity, $C_p(T)$) remained practically unchanged, whereas the zero-field transport properties (resistivity,
$\rho(T)$, and thermoelectric power, $S(T)$) showed dramatic changes. The evolution of the physical properties of
YbNi$_2$B$_2$C upon annealing was rationalized in terms of redistribution of local Kondo temperatures associated
with ligandal disorder for a small (on the order of 1\%) fraction of the ytterbium sites. The nature of the
ligandal disorder was addressed by performing single crystal X-ray diffraction and transmission electron
microscopy measurements \cite{avi04a}: lattice dislocations were found to be the dominant defect type in as-grown
single crystals. These dislocations were suggested to be responsible for the environment changes around adjacent
Yb$^{3+}$ ions leading to a distribution of the local Kondo temperatures and were shown to be (at least partially)
annealed out by heat treatment of the sample in vacuum at 950$^\circ$C. Additionally, extended-temperature-range
resistivity measurements (up to $\sim 1000$ K) \cite{avi04a} suggested that this distribution of the local Kondo
temperatures does not extend above $\sim 500-600$ K.

The aforementioned publications \cite{mia02a,avi02a,avi04a} have clearly shown that the zero-field transport
properties of YbNi$_2$B$_2$C are exceptionally sensitive to very small (effecting on the order of 1\% of the
Yb$^{3+}$ sites) perturbations of the structure. Structural defects such as these are not unique to
YbNi$_2$B$_2$C, and their effects should be taken into account if any detailed theoretical modelling of the
transport properties of materials with hybridizing moments is to be carried out.

Once YbNi$_2$B$_2$C is established as a model system for the investigation of the effects of minor structural
disorder on the physical properties of heavy fermion materials, it is of clear interest to extend the range of the
properties examined. In this work we concentrate on a very common magneto-transport characteristic of materials:
the Hall effect. It was realized quite early \cite{smi10a} that the temperature- and field- dependence of the Hall
coefficient depends on the nature and purity of a material and can be quite complex. The temperature-dependent
Hall coefficients in different heavy fermion compounds share common features (see {\it e.g.}
\cite{hur72a,chi80a,ham92a} for a review): (i) $R_H(T)$ is very large, usually it is positive for Ce- and U- based
systems and negative for Yb- based materials; (ii) with decreasing temperature $|R_H(T)|$ increases, passing
through a maximum at a temperature of the order of a coherence temperature, $T_{max} \sim T_{coh}$, then rapidly
decreasing and finally, at low temperatures (in the coherent state) approaching a constant value (or has
additional features in the case of a superconducting or magnetically ordered ground state). Since the general
behavior of the temperature-dependent Hall coefficient appears to be robust, with $T_{coh}$ as the only relevant
energy scale that is reflected as a feature in $R_H(T)$, it will be of interest to verify if in YbNi$_2$B$_2$C
annealing can modify $R_H(T)$ in a comprehensible manner.

In this work we present temperature-dependent Hall coefficient measurements on as-grown and annealed (to differing
degrees) YbNi$_2$B$_2$C single crystals and compare the results with effects of annealing on the non-magnetic,
LuNi$_2$B$_2$C, analogue.

\section{Experimental}
Single crystals of YbNi$_2$B$_2$C and LuNi$_2$B$_2$C were grown by the high-temperature flux method using Ni$_2$B
as a flux \cite{yat96a,can98a,xum94a}. Several clean, well-formed and thin YbNi$_2$B$_2$C crystals were chosen
from the same batch. Small residual droplets of flux, when present,  were polished off of the surfaces of the
crystals. The crystals were cut with a wire saw into flat bars with lengths of 3-4 mm, widths of 1.5-2.5 mm and
thicknesses of 0.15-0.2 mm. The lengths of each bar were approximately parallel to the [110] crystallographic
direction. Five contacts were attached to the samples using Pt wire with the Epotek H20E silver epoxy for
simultaneous measurements of resistivity and Hall effect (see inset to Fig. \ref{rhoTYbLu}). The temperature- and
field- dependent resistivity, $\rho(H,T)$ and the Hall resistivity, $\rho_H(H,T)$, were measured using a four
probe, ac technique ($f$ = 16 Hz, $I$ = 1-3 mA) in a Quantum Design Inc., Physical Property Measurement System
(PPMS) instrument with a separate channel assigned to each measurement. The applied magnetic field was kept
perpendicular to the electrical current for both, $\rho(H,T)$ and $\rho_H(H,T)$. To eliminate the effect of
inevitable (small) misalignment of the voltage contacts, the Hall measurements were taken for two opposite
directions of the applied field, $H$ and $-H$, and the odd component, $(\rho_H(H)-\rho_H(-H))/2$ was taken as the
Hall resistivity. After the measurements on the as-grown samples were finished the contacts were taken off from
the surface of the samples, the samples were lightly polished to remove the residues of the silver epoxy and
thoroughly cleaned with toluene and methanol, placed in a Ta foil envelope and then placed into a quartz insert of
a high vacuum annealing furnace. Similarly to the procedure described in \cite{mia02a,avi02a,avi04a}, the insert
was continuously maintained at a pressure of less then 10$^{-6}$ Torr during the annealing cycle that started with
an $\sim 1$ hour dwell at 200$^\circ$C followed by the desired time anneal at 950$^\circ$C and then a cool-down
(for 4-5 hours) to room temperature. New set of contacts was attached after annealing. Each sample was subjected
to only one annealing cycle with the measurements for as-grown material performed for each specimen used for
annealing. The same routine was used for the LuNi$_2$B$_2$C crystals for which only one annealing time, 150 hours,
was utilized.

\section{Results and Discussion}

\begin{figure}
\begin{center}
\includegraphics[angle=0,width=100mm]{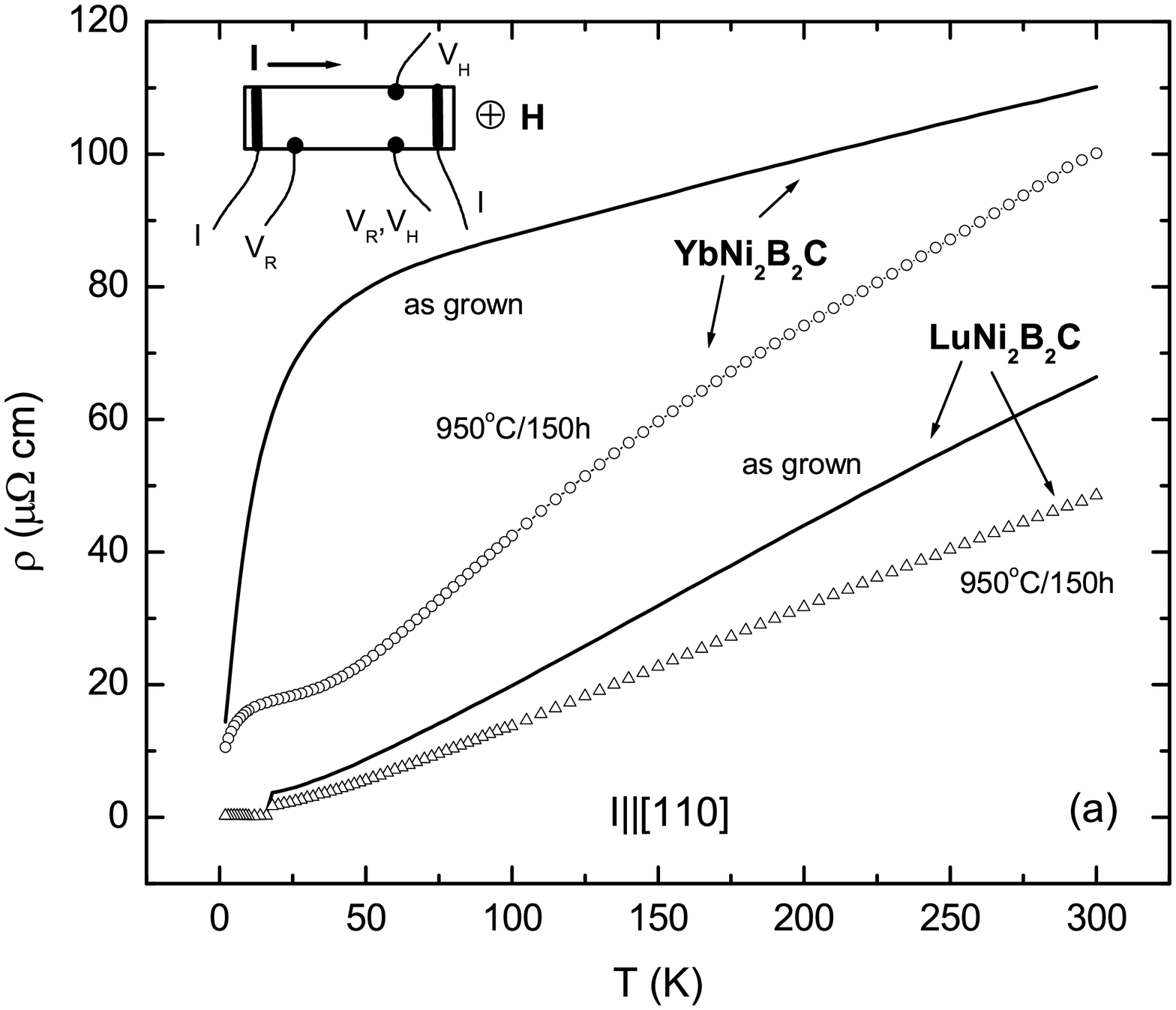}
\includegraphics[angle=0,width=100mm]{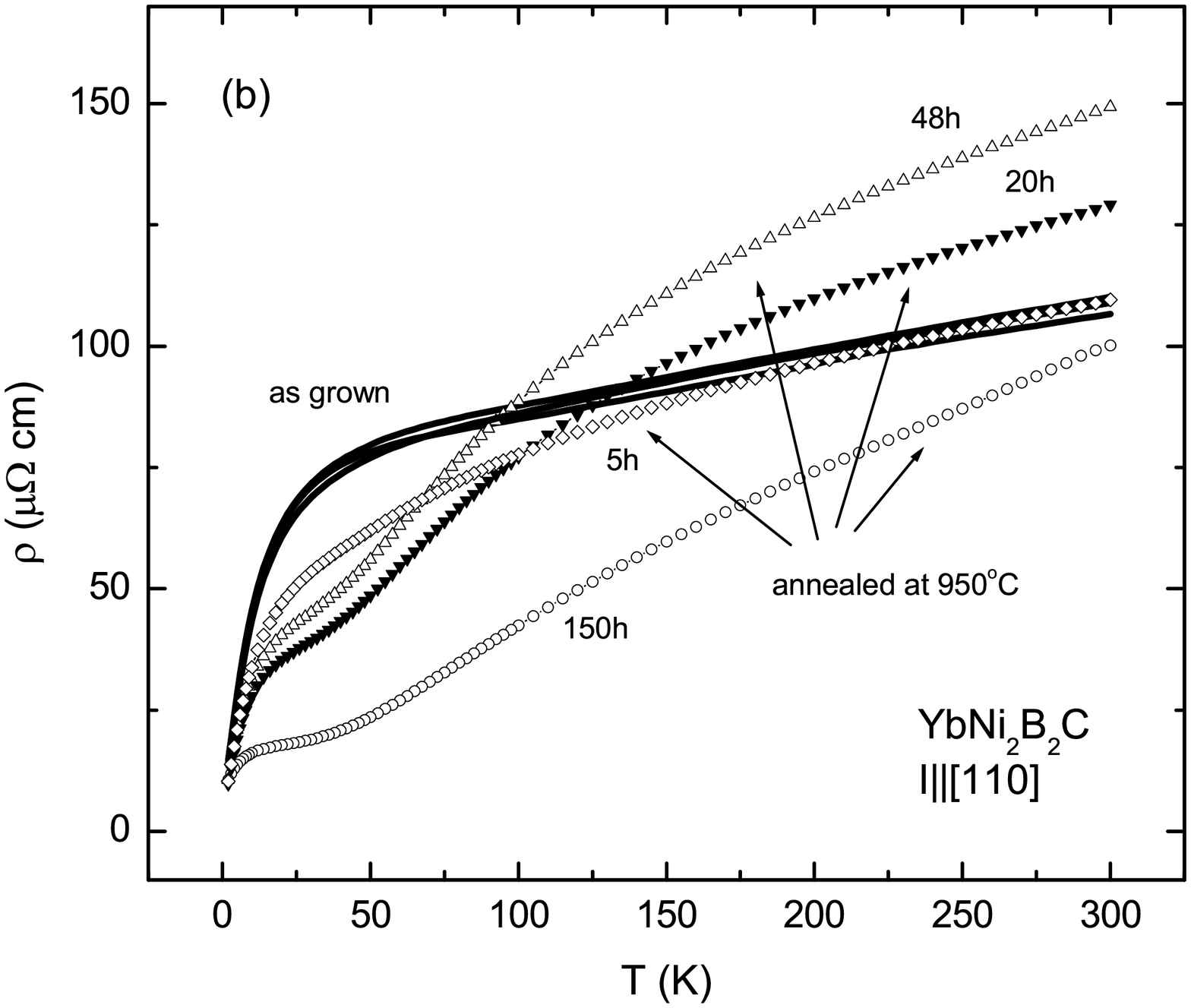}
\end{center}
\caption{(a) Zero-field resistivity of as-grown YbNi$_2$B$_2$C (line) and LuNi$_2$B$_2$C (line) and annealed at
950$^\circ$C for 150 hours YbNi$_2$B$_2$C ($\circ$) and LuNi$_2$B$_2$C ($\triangle$) samples. Inset: sketch of the
contact arrangement for simultaneous resistivity and Hall effect measurements (subscripts R and H respectively on
the sketch). (b) Zero-field resistivity of four as-grown YbNi$_2$B$_2$C crystals (lines), and the same samples
annealed at 950$^\circ$C for 150 ($\circ$), 48 ($\triangle$), 20 ($\blacktriangledown$) and 5 ($\diamond$)
hours.}\label{rhoTYbLu}
\end{figure}

The zero field resistivity data from as-grown YbNi$_2$B$_2$C and LuNi$_2$B$_2$C and annealed at 950$^\circ$C for
150 hours YbNi$_2$B$_2$C and LuNi$_2$B$_2$C samples are shown in Fig. \ref{rhoTYbLu}(a), data for different
annealing times at 950$^\circ$C are presented in Fig. \ref{rhoTYbLu}(b). The results for YbNi$_2$B$_2$C and
LuNi$_2$B$_2$C including the evolution of the $\rho(T)$ curves for the intermediate annealing times for
YbNi$_2$B$_2$C are consistent with those reported in \cite{mia02a,avi02a,avi04a} showing the high reproducibility
of the samples and the annealing procedure. All the measured as-grown YbNi$_2$B$_2$C samples (Fig.
\ref{rhoTYbLu}(b)) have very similar resistivities, further suggesting very good reproducibility within the same
batch and reasonable accuracy of the measurements of the samples' dimensions and contact positions. At first
glance the data for LuNi$_2$B$_2$C (Fig. \ref{rhoTYbLu}(a)) seem not to obey the Matthiessen's rule. We believe
that in this case the apparent difference in the $\rho(T)$ slopes between as-grown and annealed sample is caused
by somewhat higher than for the other samples (but still reasonable within the average size) accumulated error in
the dimensions and contact position measurements (a 20-25\% total correction would be required to have the
$\rho(T)$ curves for the as-grown and annealed LuNi$_2$B$_2$C data to be parallel to each other). If the room
temperature slopes of the two $\rho(T)$ plots for LuNi$_2$B$_2$C are normalized then the change associated with
annealing is the simple Matthiessen's rule shift reported in previous work \cite{mia02a}.

\begin{figure}
\begin{center}
\includegraphics[angle=0,width=120mm]{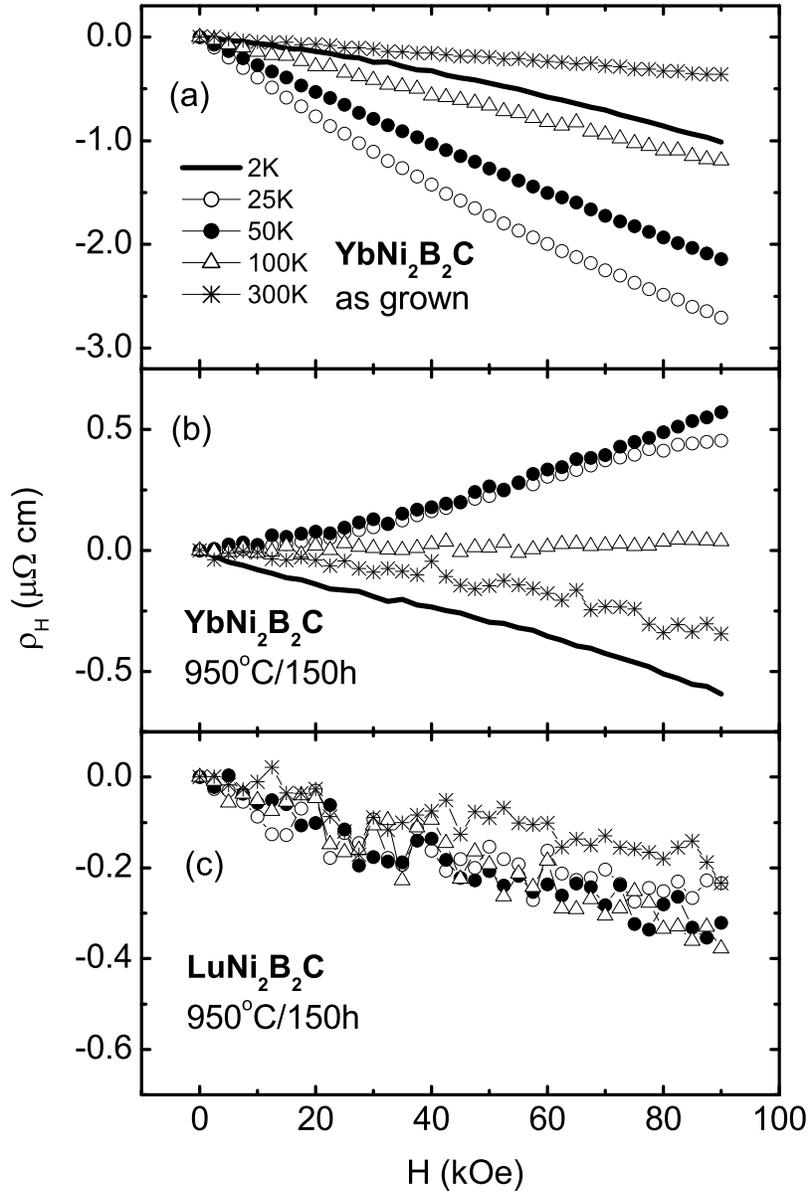}
\end{center}
\caption{Field-dependent Hall resistivity measured at different temperatures for (a) as grown YbNi$_2$B$_2$C; (b)
annealed YbNi$_2$B$_2$C and (c) annealed LuNi$_2$B$_2$C. Annealing conditions are shown on the respective panels,
symbols are consistent for all four panels.}\label{rhoHH}
\end{figure}

Fig. \ref{rhoHH} shows selected isothermal, field-dependent Hall resistivity curves for several samples. Although
for $\rho_H(H)$ taken at lower temperatures some curvature is observed, it is not large enough to significantly
change the value or the behavior of the Hall coefficient as measured in different fields. A comparison between
panels (a) and (b) clearly shows that for some temperatures the sign of the Hall resistivity is opposite for the
as-grown and the annealed YbNi$_2$B$_2$C samples, whereas at the base temperature and near the room temperature
for both samples the Hall resistivity is negative.

\begin{figure}
\begin{center}
\includegraphics[angle=0,width=120mm]{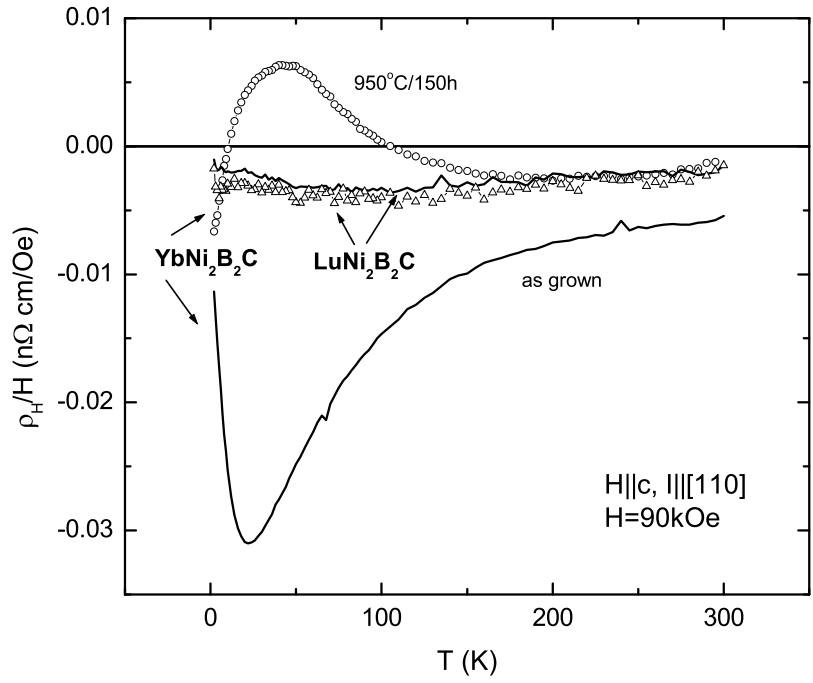}
\end{center}
\caption{The temperature-dependent Hall coefficient measured at $H = 90$ kOe for as-grown YbNi$_2$B$_2$C (line)
and LuNi$_2$B$_2$C (line) and annealed at 950$^\circ$C for 150 hours YbNi$_2$B$_2$C ($\circ$) and LuNi$_2$B$_2$C
($\triangle$) samples.}\label{RHTYbLu}
\end{figure}

The temperature-dependent Hall coefficients ($R_H(T) = \rho_H(T)/H$) of as-grown YbNi$_2$B$_2$C and LuNi$_2$B$_2$C
and annealed at 950$^\circ$C for 150 hours YbNi$_2$B$_2$C and LuNi$_2$B$_2$C are shown in Fig. \ref{RHTYbLu}. For
LuNi$_2$B$_2$C $R_H(T)$ is weakly temperature-dependent and practically unaffected by annealing; over the whole
temperature range it is negative and has values between $-(2-4) \times 10^{-12}~\Omega$ cm/Oe. These values and
general behavior are consistent with the weakly temperature-dependent Hall coefficient ($-(1-6) \times
10^{-12}~\Omega$ cm/Oe) measured on polycrystalline samples of LuNi$_2$B$_2$C and several other borocarbides
(RNi$_2$B$_2$C, R = Y, La, Gd, Ho) \cite{fis95a,nar96a,man97a,nar99a,nar99c}. The temperature-dependent Hall
coefficient of as-grown YbNi$_2$B$_2$C is negative over the whole 2-300 K temperature range, at room temperature
its value is $R_H^{300K} \approx -5 \times 10^{-12}~\Omega$ cm/Oe, this value decreases on cooling down, has a
well-defined minimum at $\approx 22$ K, with $R_H^{min}$ slightly lower than $-30 \times 10^{-12}~\Omega$ cm/Oe
and then in increases reaching $\approx -11 \times 10^{-12}~\Omega$ cm/Oe at 2 K. This behavior is consistent with
a model $R_H(T)$ behavior expected for an Yb-based heavy fermion. Curiously, the YbNi$_2$B$_2$C sample annealed at
at 950$^\circ$C for 150 hours appears to show very dissimilar behavior: its Hall coefficient is negative at room
temperature ($R_H^{300K} \approx -1.5 \times 10^{-12}~\Omega$ cm/Oe), it decreases slightly on cooling down,
passes through very shallow minimum at about 175 K, increases on further cooling, crosses zero at $\sim 100$ K,
goes through maximum at $\approx 45$ K ($R_H^{max} \approx 6.5 \times 10^{-12}~\Omega$ cm/Oe), crosses zero again
at $\sim 10$ K and reaches $\approx -7 \times 10^{-12}~\Omega$ cm/Oe at 2 K. So, as a result of annealing, a
minimum in $R_H(T)$ appears to be transformed into a maximum with its position shifted $\sim 20$ K higher. The
literature $R_H(T)$ data on a polycrystalline YbNi$_2$B$_2$C sample shown negative Hall coefficient with no minima
or maxima between 4.2 and 300 K \cite{nar99b}.

The general picture of the temperature dependence of the Hall coefficient in heavy fermion materials has been
presented in \cite{col85a,had86a,fer87a,lap87a} (see also \cite{hur72a,chi80a,ham92a} for more comprehensive
reviews). Within this picture the temperature dependence of the Hall coefficient is the result of two
contributions: a residual Hall coefficient, $R_H^{res} = \rho_H^{res}/H$, and a Hall coefficient due to the skew
scattering, $R_H^s = \rho_H^s/H$. The residual Hall coefficient is ascribed to a combination of the ordinary Hall
effect and residual skew scattering by non-hybridizing defects and impurities and, to the first approximation, is
considered to be temperature-independent, although, realistically, both the ordinary Hall effect and the residual
skew scattering may have weak temperature dependence. The temperature-dependent skew scattering contribution
($R_H^s$) at high temperatures ($T \gg T_K$, where $T_K$ is the Kondo temperature) increases as the temperature is
lowered in a manner that is mainly due to the increasing magnetic susceptibility. At lower temperatures $R_H^s$
passes through a crossover regime, then has a peak at a temperature on the order of the coherence temperature,
$T_{coh}$, and finally, on further cooling rapidly decreases (in the coherent regime of skew scattering by
fluctuations) to zero ({\it i.e.} $R_H$ ultimately levels off to the $\sim R_H^{res}$ value at very low
temperatures \cite{had86a,fer87a,lap87a}). Detailed theoretical descriptions of the temperature dependence of the
skew scattering contribution to the Hall coefficient \cite{col85a,fer87a} in the incoherent regime ($T > T_{coh}$)
offered a general expression, $R_H^s = \gamma \tilde{\chi}(T) \rho(T)$, where $\tilde{\chi}(T)$ is a reduced
susceptibility, and $\rho(T)$ is the resistivity due to the resonant scattering. In practice, $\tilde{\chi}(T) =
\chi(T)/C$, where $\chi(T)$ is a temperature dependent susceptibility and $C$ is the Curie constant and $\rho(T) =
\rho_{experimental} - \rho_{phonon} - \rho_{residual}$. In this work the resonant (magnetic) scattering
contribution to the resistivity can be approximated as $\rho_{magn} = \rho_{Yb}(T) - \rho_{Lu}(T)$, where
$\rho_{Yb}(T)$ and $\rho_{Lu}(T)$ are the experimental temperature-dependent resistivities of the YbNi$_2$B$_2$C
and LuNi$_2$B$_2$C respectively. The coefficient $\gamma$ is different for two different temperature regimes: $T
\gg T_K$ and $T_{coh} \le T \le T_K$ and depends on the details of scattering in different channels, a crossover
region exists between these two temperature regimes. This was used to explain temperature dependence of several
heavy fermion compounds (see {\it e.g.} \cite{hur72a,chi80a,ham92a}).

\begin{figure}
\begin{center}
\includegraphics[angle=0,width=120mm]{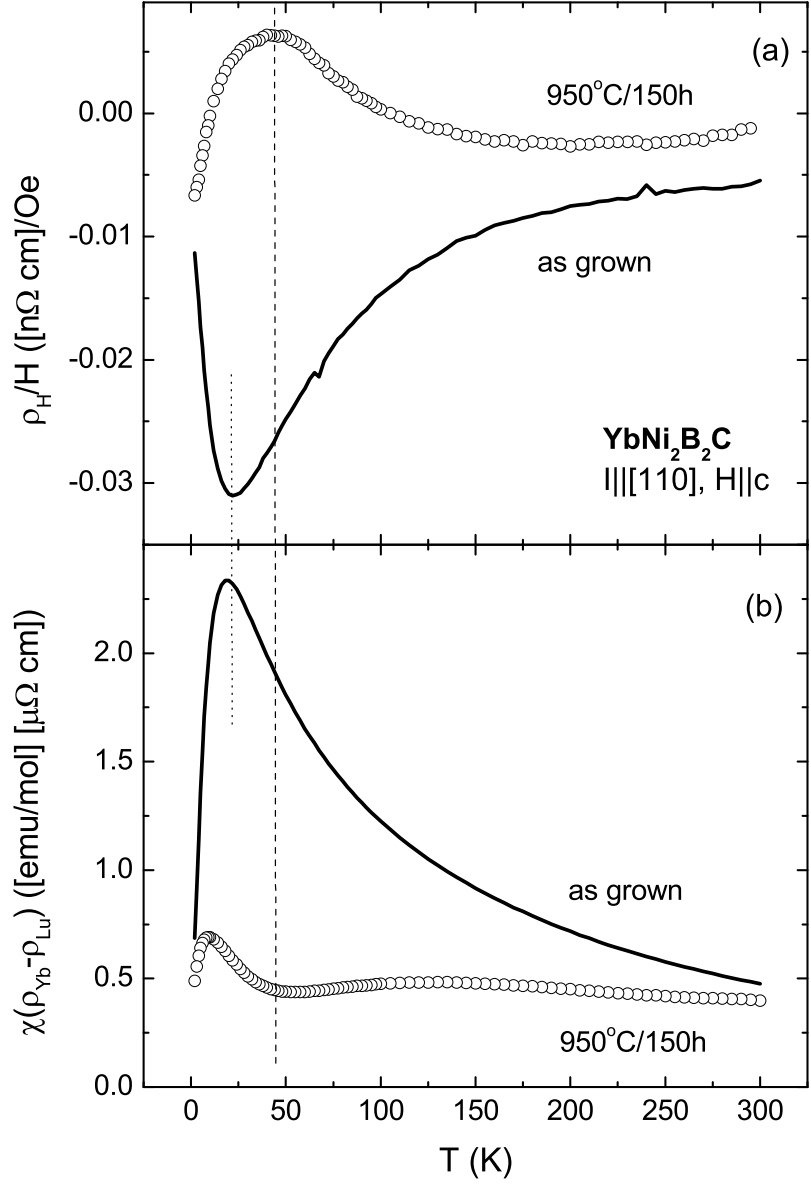}
\end{center}
\caption{Temperature-dependent (a) Hall coefficient measured at $H = 90$ kOe and (b) product of magnetic
susceptibility and resonant scattering part of resistivity, $\chi(T) (\rho_{Yb}(T) - \rho_{Lu}(T))$ for as-grown
(lines) and annealed at 950$^\circ$C for 150 hours ($\circ$) YbNi$_2$B$_2$C crystals. Dotted and dashed lines mark
positions of the extrema in the curves for as-grown and annealed samples respectively.}\label{RHchirhoYb}
\end{figure}

The temperature-dependent Hall coefficient and the product of magnetic susceptibility and the resonant scattering
contribution to resistivity [$\chi(T) (\rho_{Yb}(T) - \rho_{Lu}(T)$] for as grown and annealed at 950$^\circ$C for
150 hours YbNi$_2$B$_2$C are shown in Fig. \ref{RHchirhoYb}. For $\chi(T)$ the $H \| c$ data from \cite{avi02a}
were used ($\chi(T)$ was shown \cite{avi02a} to be annealing-independent), for the estimate of $\rho_{magn}(T)$
the experimental curves were utilized, experimental $\rho_{Lu}(T)$ data for the annealed LuNi$_2$B$_2$C sample
were utilized,  the curve was extrapolated below the superconducting transition as $\rho = \rho_0 + AT^2$. There
are striking similarities between respective $R_H(T)$ and $\chi(T)\rho_{magn}(T)$ curves. For the as-grown sample
in both curves the position of the extremum (minimum in $R_H$ and maximum in $\chi \rho_{magn}$) is approximately
the same and in the 50-300 K range both temperature dependencies are rather steep. For the annealed sample the
sign of the curvature of the $R_H(T)$ changes, the temperature dependence becomes slower, and a rather shallow
maximum at $T_{max}$ that is higher than $T_{min}$ of the as-grown sample, is observed. It appears that (around
and above $T_{coh}$) the product $\chi(T) \rho_{magn}(T)$ reproduces the main features observed in $R_H$ provided
the coefficient $\gamma$ in both regimes has a negative sign. The similarity becomes even greater if we assume
that there is a substantial, weakly temperature-dependent, possibly annealing-dependent, residual Hall effect
contribution, $R_H^{res}$ to the measured Hall coefficient. Then the apparent change of sign in the measured
temperature-dependent Hall coefficient for the annealed sample can be simply a result of an ''offset'' caused by
the $R_H^{res}$ contribution with the skew scattering contribution being negative and approximately following
$R_H^s = \gamma \tilde{\chi}(T) \rho_{magn}(T)$ (except the low temperature, coherent, region) theoretical
description.

\begin{figure}
\begin{center}
\includegraphics[angle=0,width=120mm]{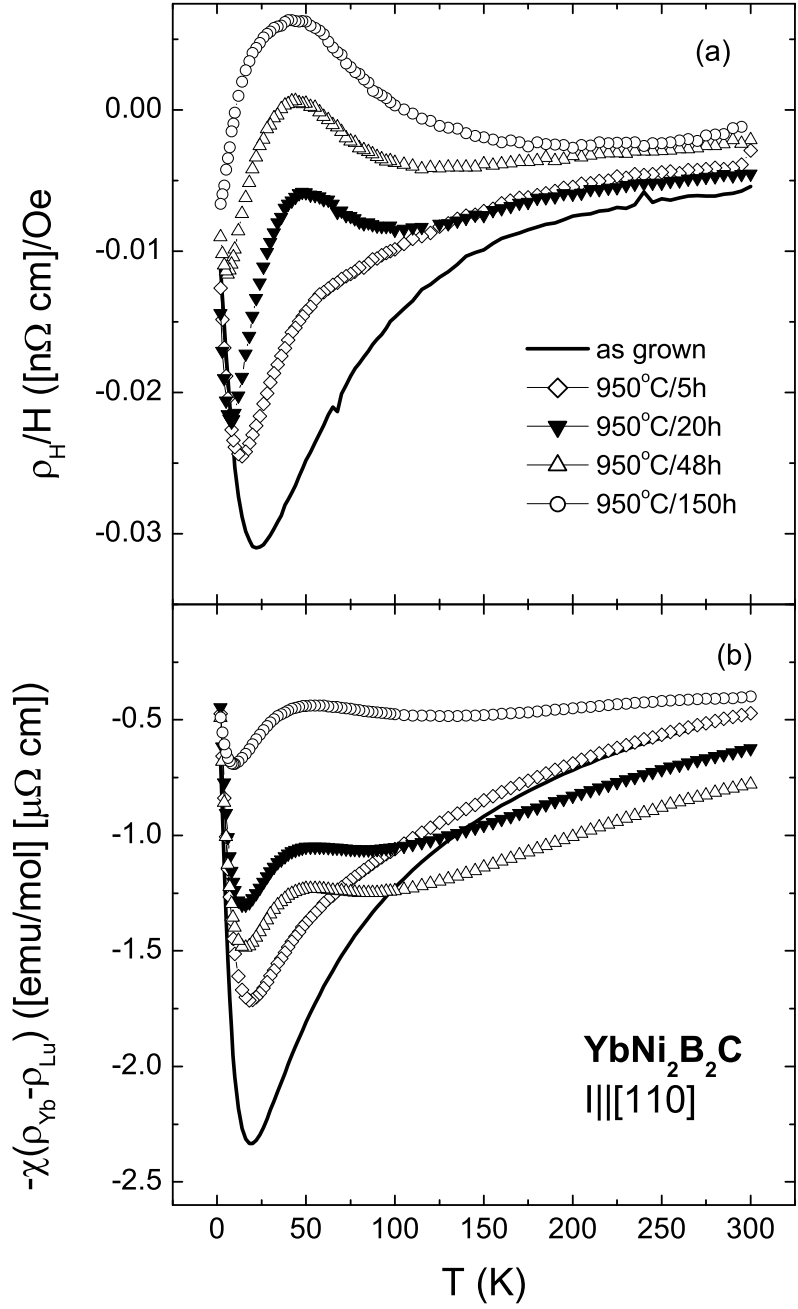}
\end{center}
\caption{Temperature-dependent (a) Hall coefficient measured at $H = 90$ kOe and (b) product of magnetic
susceptibility and resonant scattering part of resistivity, $\chi(T) (\rho_{Yb}(T) - \rho_{Lu}(T))$ for as-grown
(lines) and annealed at 950$^\circ$C for 150 ($\circ$), 48 ($\triangle$), 20 ($\blacktriangledown$) and 5
($\diamond$) hours YbNi$_2$B$_2$C crystals. Note: data in (b) are plotted as $-~\chi(T)\rho_{magn}(T)$ for easier
comparison with $\rho_H/H$ data in (a).}\label{RHchirhoYball}
\end{figure}

The same analysis can be extended to the intermediate annealing times at 950$^\circ$C. $R_H(T)$ and
($-\chi(T)\rho_{magn}(T)$) for as grown and annealed at 950$^\circ$C for 5, 20, 48, and 150 hours YbNi$_2$B$_2$C
are shown in Fig. \ref{RHchirhoYball}. Similar to the discussion above for the two extreme cases, all salient
features of the $R_H(T)$ curves (Fig. \ref{RHchirhoYball}(a)) are quite accurately reproduced in the respective
$-\chi(T) \rho(T)_{magn}$ curves which, as indicated, are plotted with negative sign, for more transparent
comparison.

The dramatic changes in $R_H(T)$ for as-grown and annealed YbNi$_2$B$_2$C can be fully understood within the
framework of the existing models of the Hall effect in heavy fermions \cite{col85a,fer87a} and our earlier
annealing studies \cite{mia02a,avi02a,avi04a}. Since the magnetic susceptibility of YbNi$_2$B$_2$C does not change
with annealing \cite{avi02a}, $R_H^s$ follows the changes in resistivity that were attributed \cite{avi02a,avi04a}
to the annealing-induced redistribution of local Kondo temperatures associated with a ligandal disorder for a
small number of ytterbium sites. No additional, Hall effect specific, mechanism seems to be required to
qualitatively understand the drastic changes in experimentally measured temperature-dependent Hall coefficient.

Whereas the temperature-dependent Hall coefficient for as-grown YbNi$_2$B$_2$C perfectly fits the general
prejudice for what should be expected for a Yb-based heavy-fermion material, we have shown that it is actually a
consequence of the ligandal disorder for a small number of the Yb-sites. Similarly to the case of zero-field
resistivity \cite{avi04a}, it should be stressed that any detailed comparison of the theoretical models with
experimental data for the Hall effect in heavy fermions should take into account this extreme sensitivity of the
Hall coefficient ({\it via} the resonant scattering part of resistivity) to the small local disorder.

\section{Summary}
We have measured the temperature dependent Hall coefficient of samples of as-grown and annealed at 950$^\circ$C
for 5-150 hours YbNi$_2$B$_2$C single crystals as well samples of as pure as-grown and annealed LuNi$_2$B$_2$C
crystals. Whereas annealing has very little effect on $R_H(T)$ of LuNi$_2$B$_2$C, the Hall coefficient of
YbNi$_2$B$_2$C changes drastically upon annealing. The changes in the measured $R_H(T)$ of YbNi$_2$B$_2$C can be
understood within a model for the skew scattering contribution to the Hall coefficient in the incoherent regime
\cite{col85a,fer87a} with some, possibly weakly temperature-dependent, residual contribution to the Hall
coefficient. The changes in $R_H(T)$ of YbNi$_2$B$_2$C upon annealing are directly connected with the changes in
zero field resistivity that were attributed to the redistribution of local Kondo temperatures for a small number
of ytterbium sites \cite{avi02a,avi04a}.

\begin{acknowledgments}
Ames Laboratory is operated for the U.S. Department of Energy by Iowa State University under Contract No.
W-7405-Eng.-82. This work was supported by the Director for Energy Research, Office of Basic Energy Sciences. We
would like to thank M. A. Avila and R. A. Ribeiro for their contribution to passionate discussions on the four
primary motivations for this and related research.
\end{acknowledgments}

\clearpage

\clearpage

\clearpage

\clearpage

\clearpage

\end{document}